# Investigating the Effect of Electrical and Thermal Transport Properties on Oxide-Based Memristors Performance and Reliability

Armin Gooran-Shoorakchaly, Sarah S. Sharif, Yaser M. Banad[*]

School of Electrical and Computer Engineering, University of Oklahoma, Norman, Oklahoma, 73019, USA
(Corresponding author's e-mail: bana@ou.edu)

**Abstract**

Achieving reliable resistive switching in oxide-based memristive devices requires precise control over conductive filament (CF) formation and behavior, yet the fundamental relationship between oxide material properties and switching uniformity remains incompletely understood. Here, we develop a comprehensive physical model to investigate how electrical and thermal conductivities influence CF dynamics in $TaO_x$-based memristors. Our simulations reveal that higher electrical conductivity promotes oxygen vacancy generation and reduces forming voltage, while higher thermal conductivity enhances heat dissipation, leading to increased forming voltage. The uniformity of resistive switching is strongly dependent on the interplay between these transport properties. We identify two distinct pathways for achieving optimal High Resistance State (HRS) uniformity with standard deviation-to-mean ratios ($\bar{\sigma}/\bar{\mu}$) as low as 0.045, each governed by different balances of electrical and thermal transport mechanisms. For the Low Resistance State (LRS), high uniformity ($\bar{\sigma}/\bar{\mu} \approx 0.009$) can be maintained when either electrical or thermal conductivity is low. The resistance ratio between HRS and LRS shows a strong dependence on these conductivities, with higher ratios observed at lower conductivity values. These findings provide essential guidelines for material selection in RRAM devices, particularly for applications demanding high reliability and uniform switching characteristics.

## Introduction

Memristive devices have emerged as promising candidates for next-generation neuromorphic computing and memory applications [1-3]. These nanoscale devices offer exceptional performance metrics, combining durability, rapid switching capabilities, and high scalability [4-6]. A particularly compelling feature is their unique ability to both store and process information within the same physical space, enabling energy-efficient computing for both in-memory and parallel processing applications [7-18].

The fundamental structure of oxide-based memristors consists of three layers: an insulating oxide layer sandwiched between top and bottom electrodes. The device operation relies on conductive filaments (CFs) that form within the oxide layer, creating paths for electrical conduction. These CFs can be repeatedly switched between high resistance states (HRS) and low resistance states (LRS) through processes of formation, breaking, and re-formation under applied voltage.

Based on their CF ionic composition, memristors fall into two categories: Electrochemical Metallization (ECM) devices, where CFs form from metal cations originating from electrochemically active electrodes [19-21], and Valence Change Memory (VCM) devices, where CFs comprise material defects, primarily oxygen vacancies ($n_D$), within metal oxide-based memristors [22-26].

VCM devices require an initial electroforming step to create a CF connecting the electrodes [27-29]. This process depends on two fundamental mechanisms: the creation of oxygen vacancies at the oxide-electrode interface through chemical reactions, and the subsequent migration of these vacancies through the bulk material [30,31], driven by both electric field forces and temperature gradients.

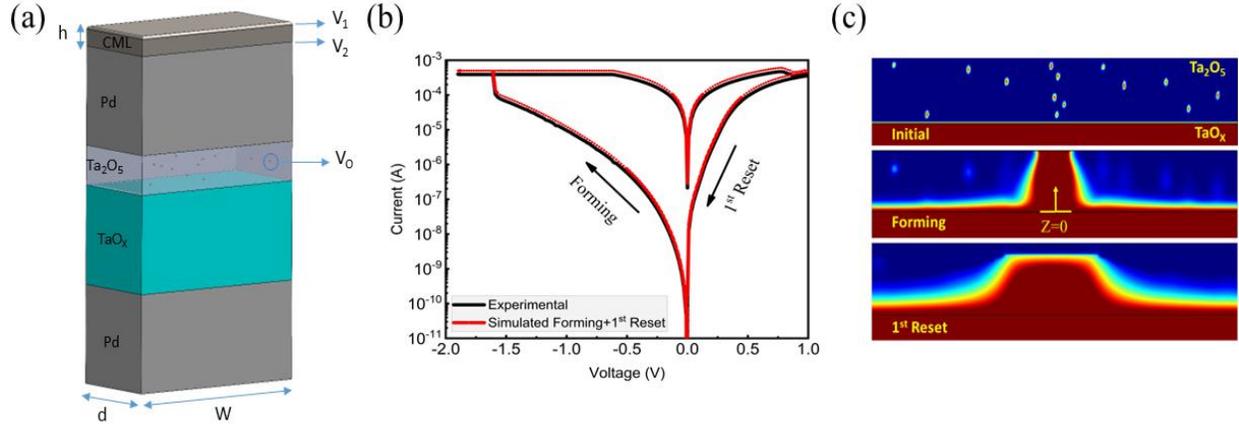

**Fig. 1.** (a) Schematic of Pd/Ta$_2$O$_5$/TaO$_x$/Pd bilayer memristor device. At the initial state, we assume a uniform doping concentration of $n_D = 1\times10^{22}$ cm$^{-3}$ within the conductive TaO$_x$ layer, which serves as the oxygen vacancy (V$_O$) reservoir. The simulation utilizes dimensions of W = 40 nm, h = 10 nm, and d = 20 nm. Here, V$_1$ represents the voltage applied to the CML while V$_2$ represents the effective voltage applied to the memristor's TE. (b) Measured and simulated dc I−V characteristics of the memristor device with $I_{CC}$ = 500 μA, illustrating the forming switching cycle. (c) 2D maps of $n_D$ were obtained in the model for forming and first reset.

Extensive research has explored various metal oxides as insulating layers in memristors, also known as resistive random-access memory (RRAM) devices. Kamiya et al. [32] demonstrated that switching processes are governed by transitions between unified and separated oxygen vacancies, with electrode materials affecting carrier injection. Clima et al. [33] investigated oxygen movement in amorphous hafnia (HfO$_x$) using molecular dynamics simulations that aligned with experimental observations. Yang et al. [34] examined electroforming mechanisms in metal oxide switches, describing how oxygen vacancies form under strong electric fields to create conductive channels. Nandi et al. [35] revealed different electroforming behaviors in Nb/NbO$_x$/Pt devices depending on NbO$_x$ film conductivity.

Despite these advances, a critical gap remains in understanding how the physical properties of metal oxides—particularly their electrical and thermal conductivity—affect CF growth during electroforming and subsequent switching operations. These properties likely play a crucial role in determining switching reliability and uniformity [36]. Previous studies on TaO$_x$-based RRAM have primarily focused on device fabrication [37], switching mechanisms [38], or basic electrical characterization [39]. For example, Lee et al. demonstrated asymmetric Ta$_2$O$_{5-x}$/TaO$_{2-x}$ bilayer structures for high endurance [4], while Wei et al. investigated redox reaction mechanisms [40]. The work by Kim et al. [41] presented a physical model for dynamic resistive switching but did not explore the fundamental relationship between material transport properties and switching uniformity. Similarly, Lee et al. [42] developed a quantitative model for TaO$_x$ memristors but focused mainly on programming dynamics rather than material property effects. While these studies have advanced our understanding of TaO$_x$-based devices, the crucial relationship between oxide transport properties (electrical and thermal conductivities) and device performance metrics (uniformity, forming voltage, and resistance ratio) remains unexplored.

In this research, we present a detailed physical model for TaO$_x$-based memristors that advances beyond existing frameworks by simultaneously considering temperature evolution and oxygen vacancy concentrations [43]. Our model addresses previous reliability prediction limitations [41] by incorporating compliance current (I$_{CC}$) to control maximum programming current during CF formation. We validate our approach using a practical 1T1R device structure, where gate voltage enables precise programming current control for better management of conductance levels and oxygen vacancy arrangements [42].

Our investigations reveal several key relationships between oxide properties and device performance. We demonstrate how electrical conductivity influences Joule heating and oxygen vacancy generation, while thermal conductivity affects heat dissipation and forming voltage requirements. These findings suggest that optimal CF uniformity can be achieved through careful selection of metal oxides with specific combinations of electrical and thermal conductivity. Following the Wiedemann-Franz law correlation between thermal and electrical conductivity in metals, we identify that transition metal oxides with reduced Lorenz numbers offer promising characteristics for RRAM applications [44]. These insights provide crucial guidelines for material selection to enhance electroforming processes and device reliability.

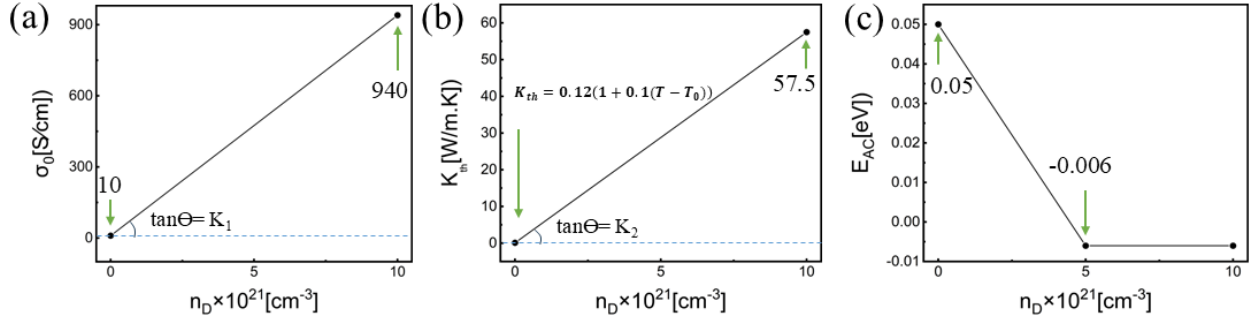

**Fig. 2**. Parameters from measurements and assumptions. (a) Electrical conductivity pre-exponential factor ($\sigma_0$), (b) assumed thermal conductivity $k_{th}$ as a function of local Vo density $n_D$, and (c) assumed activation energy for conduction ($E_{AC}$).

| Material properties | Pd | TaO$_X$ & Ta$_2$O$_5$ |
|---|---|---|
| Density ($\rho$) [Kg/m$^3$] | 11900 | 8200 |
| Heat capacity (C$_p$) [J/Kg. K] | 50 | 25 |
| Electrical conductivity ($\sigma$) [S/m] | 10$^7$ | Eq. (1) |
| Thermal conductivity (K$_{th}$) [W/M.K] | 71.8 | $K_{th} = K_{th0}(1 + \lambda(T - T_0))$ |

**Table 1**. Material parameters used in the proposed model.

| Constants | Value |
|---|---|
| Hopping distance ($a$) | 3.2Å |
| Escape attempt frequency ($f$) | 10$^{12}$ HZ |
| Linear thermal coefficient ($\lambda$) | 0.1 |
| Room temperature ($T_0$) | 293 K |
| V$_O$ diffusion barrier ($E_a$) | 0.85 eV |
| PF coefficient 1 ($\alpha$) | 5.48×10$^{-4}$ |
| PF coefficient 2 ($\beta$) | -5.7 |

**Table 2**. Material constants used in the proposed model.

## Results

### Device Structure and Physical Model

We investigate a bilayer device structure consisting of Ta$_2$O$_5$/TaO$_x$, where a high-resistance Ta$_2$O$_5$ layer is positioned above a more conductive TaO$_x$ layer (Fig. 1(a)) [45]. The structure is enclosed between palladium (Pd)-based top electrode (TE) and bottom electrode (BE). The switching mechanism primarily depends on oxygen vacancy (V$_O$) movement and redistribution between the low-resistance TaO$_x$ and high-resistance Ta$_2$O$_5$ layers [41,46]. Application of a negative voltage to the TE drives V$_O$s from the TaO$_x$ layer, which acts as a V$_O$ reservoir, into the Ta$_2$O$_5$ layer, forming a conductive filament (CF) that connects the TE to the conductive TaO$_x$ layer. This set process transitions the device to a low-resistance state. Conversely, applying a positive voltage during the reset process repels V$_O$s from the Ta$_2$O$_5$ layer, rupturing the conductive filament and returning the device to a high resistance state. Figure 1(b) demonstrates the measured and simulated DC I-V characteristics during the forming process with a compliance current (I$_{CC}$) of 500 μA, while Figure 1(c) shows the two-dimensional maps of V$_O$ concentration (n$_D$) during the forming process. Our simulation results show excellent agreement with experimental measurements [42].

In our analysis, we assume a uniform V$_O$ concentration of n$_D$=1×10$^{22}$ cm$^{-3}$ and an electrical conductivity of 10$^5$ Sm$^{-1}$ within the conductive TaO$_x$ layer, based on density functional theory calculations and experimentally measured conductivity values for TaO$_x$ films [40]. These V$_O$ defects are assumed to exhibit the same local conductivity (10$^5$ Sm$^{-1}$) as the TaO$_x$ layer. Conversely, the stoichiometric Ta$_2$O$_5$ film is modeled with a significantly lower V$_O$ concentration (n$_D$ =1×10$^{16}$ cm$^{-3}$), making it highly resistive. In this framework, the electrical conductivity ($\sigma$) of the oxide material is expressed as a function of n$_D$, temperature (T), and electric field (E). The relationship is described by the equation:

$$\sigma_{Ta_2O_{5-x}} = \sigma_0 \, exp\left(-\frac{E_{AC}}{KT}\right) + \sigma_{PF}(E, T) \qquad (1)$$

where σ₀ is a prefactor, $E_{ac}$ is the activation energy for electron conduction, and $\sigma_{PF}(E,T) = exp(\frac{293}{T}) \cdot (a\sqrt{E} + \beta)$ represents the Poole−Frenkel (PF) conduction term. To simulate the set and reset transitions, we solve three partial differential equations (PDEs) self-consistently: (1) a continuity equation for the drift and diffusion of $V_O$, (2) a current continuity equation for electrical conduction, and (3) a Fourier equation to account for Joule heating. The drift and diffusion of $V_O$ migration are modeled using a simplified one-dimensional rigid point ion framework originally introduced by Mott and Gurney [47]. Although Mott and Gurney's ion-hopping model was developed for oxygen ions, we extend its applicability to oxygen vacancies, with adjustments made to specific physical parameters. Based on Eq.(5), the diffusion and the drift velocity are described by $D = \frac{1}{2}a^2 f exp(-\frac{E_a}{KT})$, $v = af exp(-\frac{E_a}{KT}) sinh(-\frac{qaE}{KT})$, respectively, where $f$ is the attempt-to-escape frequency ($10^{12}$ Hz) [41], $a$ is the effective hopping distance (0.32 nm), and $E_a$ is the activation energy required for $V_O$ migration (0.85 eV). In reality, the effective hopping distance can vary within a range of a few nanometers (typically <1 nm). However, the value of 0.32 nm is used in this study to best fit the experimental data. This discrepancy likely arises from the different physical properties of oxygen vacancies compared to oxygen ions. Incorporating both drift and diffusion, the time-dependent change in $V_O$ concentration ($n_D$) is described by the following continuity equation:

$$\frac{\partial n_d}{\partial t} = \nabla \cdot (D\nabla n_d - v n_d + DSn_d \nabla T) \qquad (2)$$

where $D\nabla n_D$ and $vn_D$ are the Fick diffusion flux and the drift flux terms, respectively [41]. The $DSn_D\nabla T$ term corresponds to the Soret diffusion flux, where S is the Soret coefficient ($S = -E_a/kT^2$). Soret diffusion describes the movement of particles along a temperature gradient, which contributes to the formation of stable conductive filaments (CFs) under high-temperature conditions. In the simulation, $V_O$s migrate toward areas of higher temperature, particularly within the filament region, when exposed to a temperature gradient. This migration occurs due to the enhanced diffusivity of oxygen ions at elevated temperatures. As a result, the process facilitates the development of a stable CF composed of $V_O$ particles, even in high-temperature environments. To model this behavior, Eq. (2), which governs Soret diffusion, is solved in conjunction with Eq. (3), the continuity equation for electrical conduction, and Eq. (4), the Fourier equation for Joule heating. This self-consistent approach follows the methodology proposed by Ielmini et al. [48].

$$\nabla \cdot \sigma \nabla \Psi = 0 \qquad (3)$$

$$\rho C_p \frac{\partial T}{\partial t} - \nabla \cdot k_{th} \nabla T = J \cdot E \qquad (4)$$

These equations are solved simultaneously using numerical methods (COMSOL) to calculate electrical conductivity (σ), electric potential (Ψ), oxygen vacancy concentration ($n_D$), and temperature (T). The simulated structure has a width of 40 nm, a depth of 20 nm, and consists of layers with the following thicknesses: a 35 nm palladium bottom electrode (BE), a 30 nm $V_O$ reservoir layer ($TaO_x$) a 5 nm switching layer ($Ta_2O_5$), and a 50 nm palladium top electrode (TE). These dimensions and layer compositions are designed to replicate the physical properties of the system under study. Parameter definitions and values are detailed in Tables 1 and 2 and illustrated in Fig. 2, as referenced in prior works [48,49].

To solve Eqs. (3) and (4), models for both electrical conductivity (σ) and thermal conductivity ($k_{th}$) are essential. We consider these properties to be dependent on $n_D$, as illustrated in Fig. 2(a) and 2(b). Following Eq. (5), σ₀ increases linearly from 10 to 940 $Ω^{-1}$ $cm^{-1}$ with increasing $n_D$. Fig. 2(c) presents the conduction activation energy $E_{AC}$ used in our calculations, which is -0.006 eV at high $n_D$ and increases linearly to 0.05 eV with decreasing $n_D$, corresponding to measured data for LRS and HRS, respectively.

Furthermore, a linear relationship between $k_{th}$ and $n_D$ is assumed, as shown in Fig. 2(b). The minimum value for $n_D=0$ is associated with the thermal conductivity of insulating $Ta_2O_5$, $K_{Ta_2O_5} = 0.12$ $Wm^{-1}K^{-1}$ at $T_0 = 300$ K. Additionally, the thermal conductivity $K_{Ta_2O_5}$ is assumed to have a linear temperature dependence, expressed as:

$$K_{Ta_2O_5} = 0.12(1+\lambda(T-T0) \qquad (5)$$

where λ= 0.1 is the linear thermal coefficient. The maximum $k_{th}$ value at high $n_D$ is equivalent to that of metallic CF, represented by the thermal conductivity of tantalum, $k_{Ta} = 57.5$ W $m^{-1}$ $K^{-1}$.

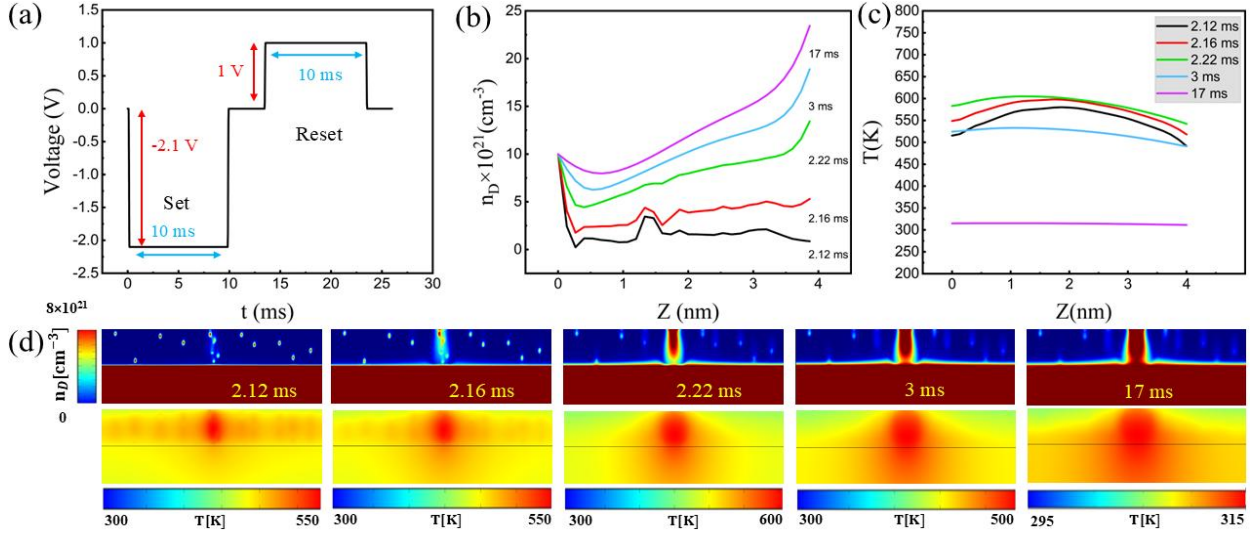

**Fig. 3** Forming process. (a) Schematic of the device during pulse forming. (b) Evolution of the oxygen vacancy concentration ($n_D$) across the $Ta_2O_5$ layer thickness (Z) at different times. (c) Temperature distribution across the Z over time. (d) 2D maps of oxygen vacancy concentration and temperature at different times during the forming process.

For boundary conditions, we maintain a constant temperature (300 K) at the surfaces of BE, TE, and the current modulation layer (CML). The BE remains grounded throughout the simulation, while voltage is applied to the CML's top surface. To implement $I_{CC}$ during forming and set processes, we define the CML conductance through Eqs. (6), (7), and (8):

$$\sigma_{CML} = \frac{I_{CC} \cdot h}{|V_1 - V_2| \cdot w \cdot d} \quad \text{if } E_{CML} \geq E_{max} \quad (6)$$

$$\sigma_{CML} = I_{CC\_Sigma} \quad \text{if } E_{CML} \leq E_{max} \quad (7)$$

$$E_{max} = \frac{I_{CC}}{I_{CC\_Sigma} \cdot w \cdot d} \quad (8)$$

Here, $V_1$ denotes the voltage applied to the top surface of the CML, while $V_2$ represents the effective voltage at the memristor's TE, as depicted in Fig.1(a). When the electrical field applied to the CML is below $E_{max}$, the CML conductance maintains $I_{CC\_Sigma}$ ($10^5$ Sm$^{-1}$). For fields exceeding $E_{max}$, the conductance follows Eq. (8). During Reset process, $\sigma_{CML}$ remains constant at $I_{CC\_Sigma} = 10^5$ Sm$^{-1}$.

## The Conductive Filament (CF) Growth Behavior During the Electroforming Process

Figure 3 illustrates the detailed dynamics of CF formation in our device. Fig.3(a) depicts the applied voltage pulses during forming and reset operations, with fixed amplitudes of -2.1 V for the set (forming) and 1 V for reset, each lasting 10 ms. During the forming process, the high-resistance $Ta_2O_5$ layer initially allows minimal current flow due to its insulating properties. Before CF formation, Poole–Frenkel (PF) emission dominates conduction through defect-free regions of the $Ta_2O_5$ layer.

The evolution of oxygen vacancy concentration ($n_D$) across the device thickness (z) over time (2.12 ms to 17 ms) is shown in Fig. 3(b). We observe a progressive increase in $n_D$, particularly near the $Ta_2O_5$/TE interface. This increase is driven by the applied electric field, which generates localized heating and promotes $V_O$ migration toward the TE, facilitating CF formation within the oxide layer.

Figure 3(c) presents the temporal evolution of temperature distribution across the device. As the forming pulse continues, Joule heating leads to significant temperature elevation, particularly near the TE where filament formation occurs. Local temperatures exceed 500 K in specific regions, further facilitating $V_O$ migration. Upon voltage removal, temperature decreases, marking the completion of the forming process.

The 2D maps in Fig. 3(d) provide spatial visualization of both $n_D$ and temperature distributions at various time points during forming. These maps demonstrate the progressive concentration of heat and $V_O$ migration in specific

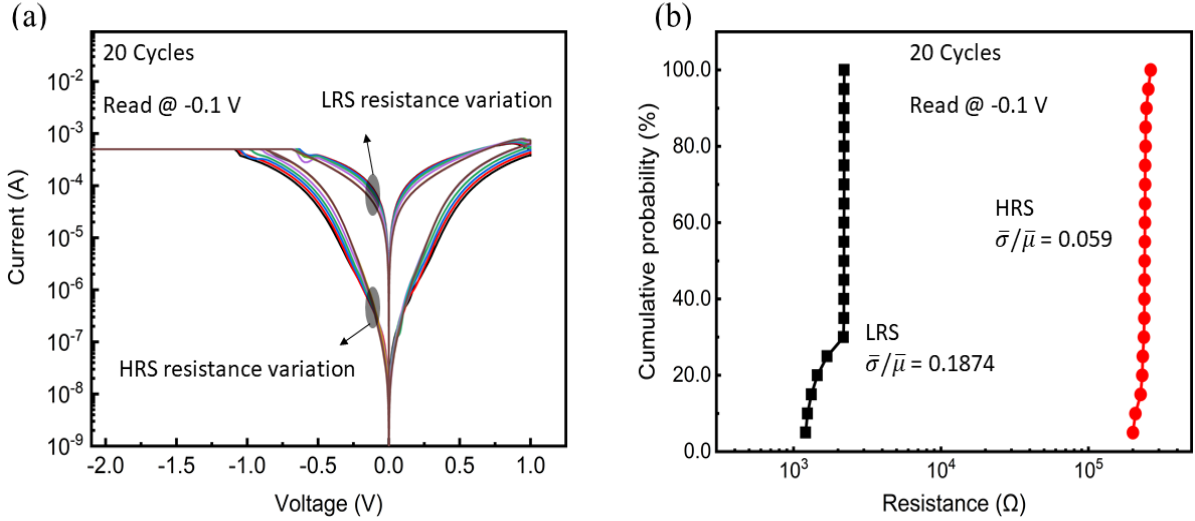

**Fig. 4** (a) Current–voltage curves of all 20 consecutive switching cycles. (b) Cumulative cycle-to-cycle resistance distribution of Pd/ $Ta_2O_5$/$TaO_x$ /Pd RRAM device. Both LRS and HRS show resistance variability. The uniformity of resistance states is quantified using the standard deviation-to-mean ratio ($\bar{\sigma}/\bar{\mu}$), with HRS showing $\bar{\sigma}/\bar{\mu}$= 0.059 and LRS exhibiting $\bar{\sigma}/\bar{\mu}$= 0.1874. The LRS and HRS are measured at V = -0.1V.

regions, leading to controlled CF formation. The spatial and temporal evolution of these parameters reveals the complex interplay between electrical and thermal effects during the forming process.

**Resistive Switching Mechanism**

While RRAM technology shows promising memory characteristics, resistance variability and incomplete understanding of switching mechanisms remain critical challenges for practical applications. Both LRS and HRS exhibit resistance variability, manifesting as temporal and spatial fluctuations that affect the stability of the memory window.

Figure 4(a) presents I-V curves from 20 consecutive DC switching cycles, demonstrating resistance variations in both LRS and HRS, reducing the memory window, and thereby the HRS to LRS (HRS/LRS) ratio. As the memory window shrinks, the device becomes less robust and more susceptible to noise and errors during read/write operations. Notably, cycles 7-20 show minimal variation in hysteresis curve uniformity, resulting in overlapping characteristics.

Figure 4(b) provides cumulative probability distributions of resistance states, with standard deviation-to-mean ratio ($\bar{\sigma}/\bar{\mu}$) of 0.059 for HRS and 0.1874 for LRS. Higher LRS variability and overlapping resistance regions highlight the challenges in achieving consistent device performance. These findings emphasize the need for a deeper understanding of the mechanisms governing filament formation and rupture, along with advancements in material engineering and device optimization to minimize such variations.

**Effect of Oxide Properties on the CF Growth Behavior**

Our simulation results reveal that the local electric field and temperature are key factors influencing the movement of oxygen vacancies and the subsequent growth of CF. Following Eqs. (3) and (4), the electrical and thermal conductivity properties of host metal oxides, significantly influence these processes. Given that σ and kth values for metal oxides are substantially lower than their metallic states, we simplify our analysis by assuming uniform initial σ and $k_{th}$ for various pristine oxides (such as $Ta_2O_5$, $HfO_2$, etc). These properties are linearly extrapolated to their metallic state values as $n_D$ increases, with the slopes defined by parameters $K_1$ and $K_2$.

In metal oxide-based RRAM devices, oxygen vacancies act as n-type donors, creating distinct behavioral regimes. For regions with lower oxygen vacancy density ($n_D < 5\times10^{27}$ $m^{-3}$), semiconductor-like behavior dominates, with EAC increasing linearly from 0.0 eV to 0.05 eV as $n_D$ decreases, matching reported values for $TaO_x$ reset states. At higher concentrations ($n_D > 5\times10^{27}$ $m^{-3}$), EAC becomes 0.0 eV, indicating Fermi level pinning in the conduction band - characteristic of the CF's metallic nature. The relationship between kth and $n_D$ follows the Wiedemann-Franz Law

[48,50], providing theoretical justification for our linear approximation approach. This framework enables systematic analysis of how electrical and thermal properties influence CF growth through the tuning of $K_1$ and $K_2$ parameters.

Figure 5 demonstrates electrical conductivity's effect on CF growth. The total current evolution with different electrical conductivities ($K_1$ = 6.2, 9.5, and 18.8 Sm$^{-1}$) shows that while initial current patterns remain similar, the forming voltage ($V_f$) decreases monotonically with increasing $K_1$. The 2D maps in Fig. 5(c) and corresponding 1D profiles in Fig. 5(b) reveal that decreasing $K_1$ leads to significant $n_D$ accumulation near the TE/Ta$_2$O$_5$ interface, resulting from increased inhomogeneity in dopant density distribution at lower electrical conductivity values. As $K_1$ reaches 18.8 Sm$^{-1}$, the $n_D$ becomes more homogenous and the CF becomes uniform in morphology.

The 1D electric field distributions near the Ta$_2$O$_5$/TaO$_x$ interface (y= 0 and z= 0 nm), shown in Fig. 5(d), provide crucial insights into the forming process. At low electrical conductivity ($K_1$ = 6.2 Sm$^{-1}$), the reduced Joule heating effect leads to lower local temperatures (Fig. 5(e), pink curve). This necessitates a higher forming voltage, which in turn creates a stronger local electric field (Fig. 5(d), pink curve) that accelerates $n_D$ migration toward the TE. As electrical conductivity increases ($K_1$ = 9.4 Sm$^{-1}$), enhanced heat generation reduces the forming voltage, resulting in a smaller local electric field enhancement (Fig. 5(d), red curve) and decreased driving force for $n_D$ migration. At the highest conductivity ($K_1$ = 18.8 Sm$^{-1}$), extensive heat generation leads to the lowest electric field (Fig. 5(d), blue curve), significantly reducing $n_D$ drift toward the TE. This interplay between electrical conductivity, local electric field, and temperature demonstrates how these parameters collectively influence CF formation dynamics. Higher electrical conductivity promotes more uniform CF formation through reduced electric field gradients, while lower conductivity leads to more localized CF formation due to stronger field enhancement effects.

Fig. 6 illustrates the effect of $k_{th}$ (by slope $K_2$) on the CF growth. Here, we apply the same boundary conditions of electrical bias as the electrical conductivity case. Fig. 6(a) shows the temporal evolution of total current with different $k_{th}$ (represented as $K_2$ = 2.5, 5.75, and 11.5 Wm$^{-1}$K$^{-1}$). The lower $k_{th}$ ($K_2$ = 2.5 Wm$^{-1}$K$^{-1}$) results in a higher temperature (pink curve in Fig. 6(e)) which promotes the $n_D$ generation at a reduced forming voltage of -1.62 V (Fig. 6(a)). In this case, the induced $n_D$ migrates to the TE and the drift flux from TaO$_x$ to Ta$_2$O$_5$ is suppressed due to the decreasing local electrical field (pink curve in Fig. 6d). As the $k_{th}$ ($K_2$) increases, the heat generated by the Joule heating effect can be easily dissipated and the local temperature decreases (red curve in Fig. 6(e)), which inhibits the $n_D$ generation and increases the CF forming voltage up to -1.78 V (Fig. 6(a)). When $k_{th}$ further increases ($K_2$ = 11.5 Wm$^{-1}$K$^{-1}$), the induced $n_D$ migration from TaO$_x$ to Ta$_2$O$_5$ becomes much easier under a higher local electrical field (blue curve in Fig. 6d). Thus, more $n_D$ accumulation near the TE and a larger $n_D$ width near the TE are observed (Fig. 6(c) and blue curve in Fig. 6(b)).

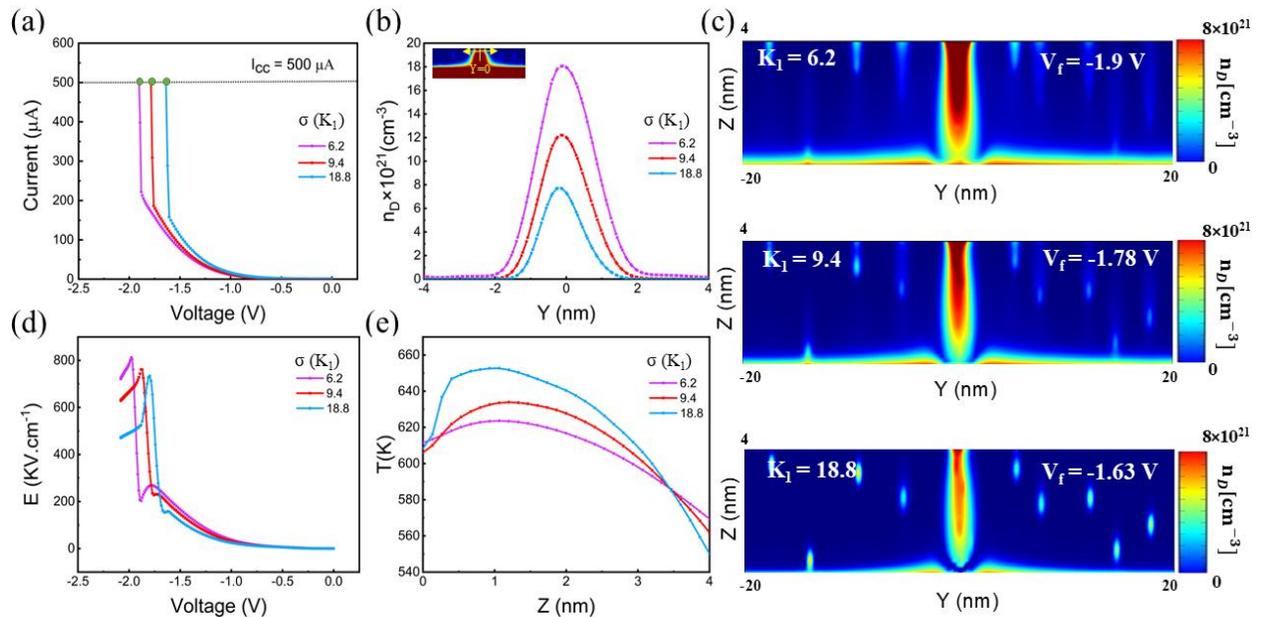

**Fig. 5** Effect of electrical conductivity on the CF growth. (a) Current characteristic with different electrical conductivity (by $K_1$). (b) Horizontal 1D $n_D$ profiles for different $K_1$ at final state. The y = 0 position is the center of the filament. (c) 2D maps of distributions of $n_D$ at final state. (d) Calculated 1D profiles of local electric field (E) near the Ta$_2$O$_5$/TaO$_x$ interface (y= 0 and z= 0 nm). (e) temperature (T) along the center of CF (y = 0, z = 0–4 nm) at final state.

Figure 7 provides a comprehensive analysis of how $K_1$ and $K_2$ influence the forming voltage and the resistance ratio ($R_{HRS}/R_{LRS}$). Figure 7(a) presents the forming voltage as a function of $K_1$ and $K_2$, while Figure 7(b) shows the corresponding resistance ratio under similar conditions. From Figure 7(a), it is evident that a low electrical conductivity needs a stronger forming voltage (higher negative value), especially at higher thermal conductivities. forming voltage decreases with increasing electrical conductivity and decreasing thermal conductivity. This trend can be attributed to the Joule heating effect: increased electrical conductivity generates more localized heat, facilitating oxygen vacancy generation and reducing the energy required for CF formation. Conversely, higher thermal conductivity enhances heat dissipation, suppressing vacancy generation and necessitating an elevated forming voltage. This relationship highlights the critical balance between electrical and thermal properties for achieving efficient forming processes.

Figure 7(b) demonstrates that the $R_{HRS}/R_{LRS}$ ratio, a crucial performance metric for RRAM devices, strongly depends on these parameters. The resistance ratio shows distinct regions of behavior across the $K_1$-$K_2$ parameter space. A high $R_{HRS}/R_{LRS}$ ratio (>250) is observed in the region where 11 $Sm^{-1} < K_1 < 18$ $Sm^{-1}$ and 5 $Wm^{-1}K^{-1} < K_2 < 6$ $Wm^{-1}K^{-1}$, corresponding to moderate forming voltages (-1.75V to -1.65V). This optimal region can be explained by two key physical mechanisms. In this regime, the moderate electrical conductivity ($K_1$) provides sufficient current flow to form stable conductive filaments during SET operation while allowing effective rupture during RESET. The balanced electrical transport prevents excessive Joule heating that could lead to uncontrolled filament growth or incomplete rupture. Additionally, the relatively low thermal conductivity ($K_2$) in this region helps maintain localized heating during switching operations. This localization is crucial for achieving high $R_{HRS}/R_{LRS}$ ratios as it enables complete filament rupture during RESET, leading to higher HRS resistance, while ensuring controlled filament formation during SET, maintaining stable LRS resistance. Conversely, lower resistance ratios are observed in regions with lower electrical conductivity and higher thermal conductivity. This reduction occurs because lower electrical conductivity limits the completeness of filament formation during SET, resulting in higher LRS resistance. Simultaneously, higher thermal conductivity leads to rapid heat dissipation, preventing efficient local heating needed

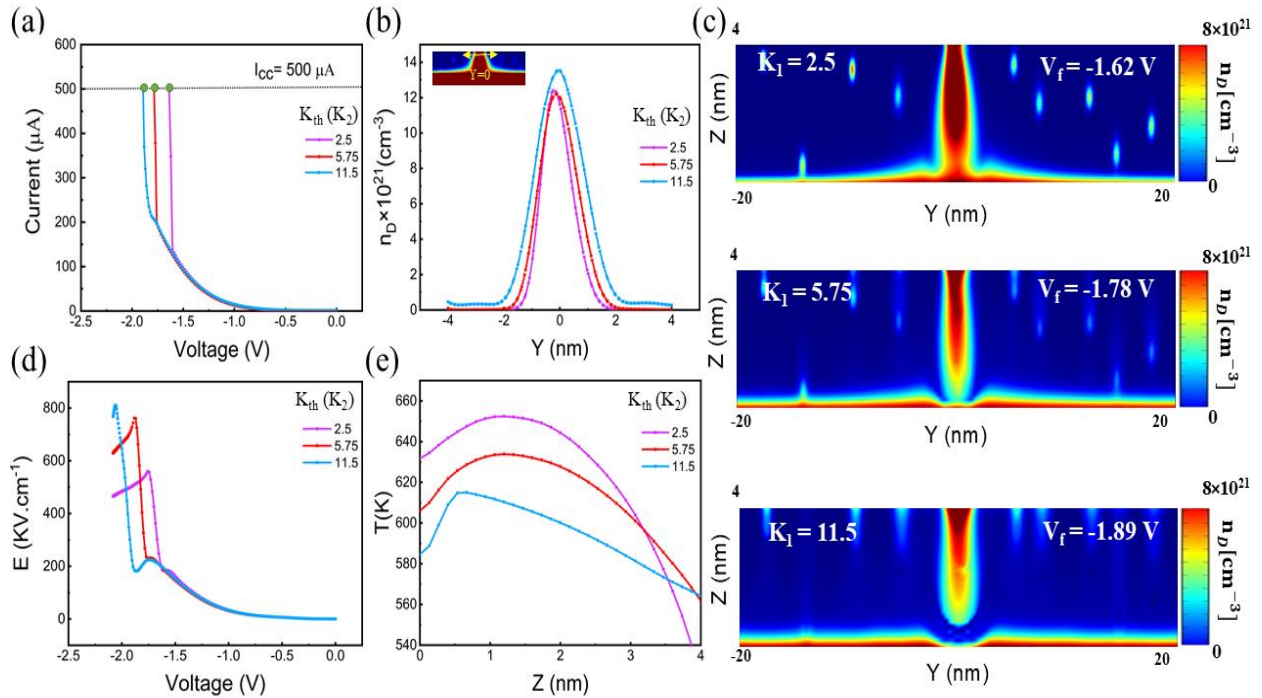

**Fig. 6** Effect of thermal conductivity on the CF growth. (a) Current characteristic with different thermal conductivity (by $K_2$). (b) Horizontal 1D $n_D$ profiles for different $K_2$ at final state. The y = 0 position is the center of the filament. (c) 2D maps of distributions of $n_D$ at final state. (d) Calculated 1D profiles of local electric field (E) near the $Ta_2O_5/TaO_x$ interface (y= 0 and z= 0 nm). (e) temperature (T) along the center of CF (y = 0, z = 0–4 nm) at final state.

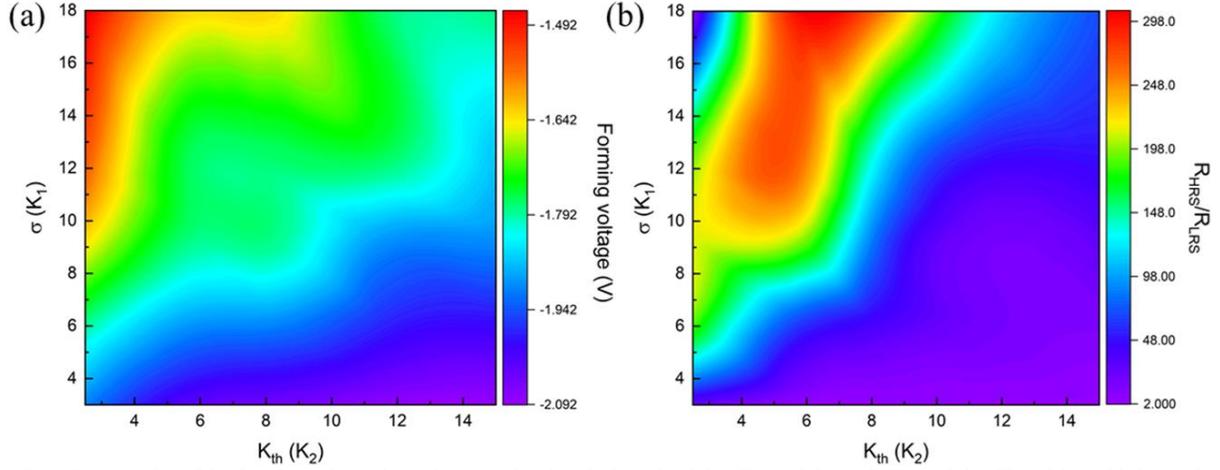

**Fig. 7** (a) Contour plot of the forming voltage dependence on the electrical conductivity ($K_1$) and thermal conductivity ($K_2$) of the oxide layer. (b) Contour plot illustrating the ratio of high resistance state (HRS) to low resistance state (LRS), $R_{HRS}/R_{LRS}$, as a function of $K_1$ and $K_2$.

for complete filament rupture during RESET, resulting in lower HRS resistance. These results indicate that optimal $R_{HRS}/R_{LRS}$ ratios require careful balancing of Joule heating (controlled by $K_1$) and heat dissipation (controlled by $K_2$) to achieve both efficient filament formation and complete rupture during switching operations. This balance is critical for maintaining the large memory window necessary for reliable and energy-efficient device resistive switching operation.

The graphs in Fig. 8 illustrate the uniformity analysis of resistive switching characteristics, mapping the standard deviation ($\bar{\sigma}$)-to-mean ($\bar{\mu}$) ratio as a function of electrical conductivity ($K_1$) and thermal conductivity ($K_2$). This analysis is crucial for understanding device reliability and optimization [51]. Figure 8(a) shows the HRS uniformity map with two distinct regions of optimal performance. The first occurs at mid-range $K_1$ ($\approx 8.2$ Sm$^{-1}$) and low $K_2$ ($\approx 5.5$ Wm$^{-1}$K$^{-1}$), achieving excellent uniformity ($\bar{\sigma}/\bar{\mu} \approx 0.045$). In the first region, the moderate electrical conductivity combined with low thermal conductivity creates ideal conditions for filament rupture. The moderate electrical conductivity prevents excessive current flow that could cause irregular rupture patterns, while the low thermal conductivity ensures heat remains localized. This localized heating leads to consistent temperature profiles during RESET, resulting in reproducible filament rupture locations and dimensions. In the second region ($K_1 \approx 12$ Sm$^{-1}$, $K_2 \approx 9.5$ Wm$^{-1}$K$^{-1}$), the physics is different. Higher electrical conductivity enables more efficient oxygen vacancy movement due to stronger electric fields, while moderate thermal conductivity provides balanced heat distribution.

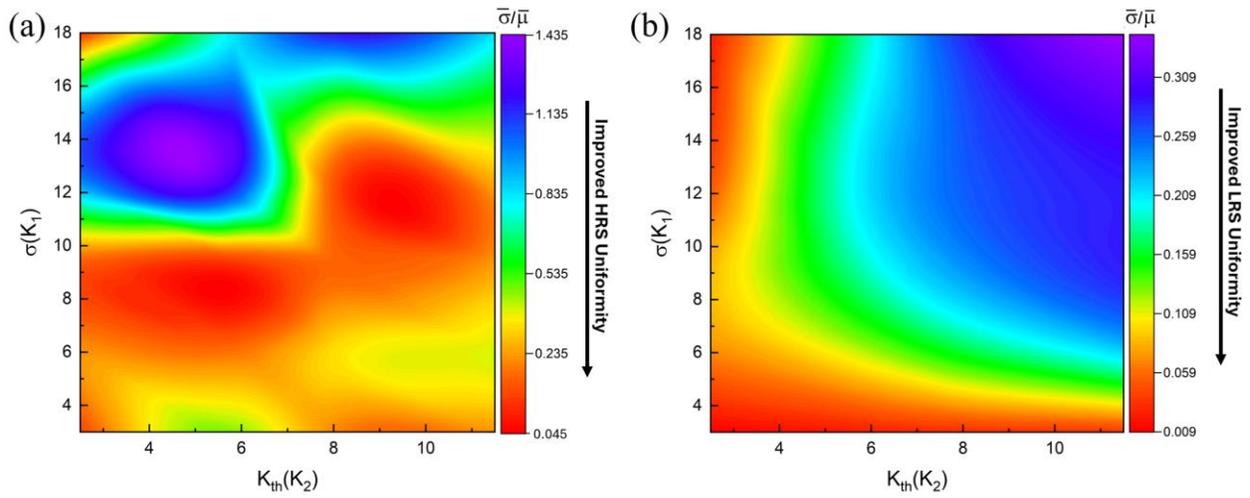

**Fig. 8** Contour plots illustrating the standard deviation-to-mean ratio ($\bar{\sigma}/\bar{\mu}$) for (a) HRS and (b) LRS as a function of electrical conductivity ($K_1$) and thermal conductivity ($K_2$).

This combination allows oxygen vacancies to migrate more uniformly during RESET, leading to consistent filament rupture despite the higher thermal dissipation.

Figure 8(b) reveals that LRS uniformity has a simpler dependency on electrical conductivity and thermal conductivity, exhibiting the optimal uniformity of $\bar{\sigma}/\bar{\mu} \approx 0.009$. For LRS uniformity, the physics is governed by filament formation dynamics. When either $K_1$ or $K_2$ is low, or both, the SET process becomes more controlled as low thermal conductivity concentrates Joule heating, ensuring consistent filament formation locations and low electrical conductivity limits current flow, preventing random branching of filaments during formation. However, when both conductivities are high, excess current flow combined with widespread heat distribution leads to less controlled filament formation, reducing uniformity. The interesting case of high LRS uniformity at high $K_1$ ($\approx 14$ Sm$^{-1}$) and low $K_2$ actually results in poor HRS uniformity because the same conditions that enable consistent filament formation (high electrical conductivity) make controlled rupture more difficult, highlighting the inherent trade-off between SET and RESET processes in these devices.

The relationship between these parameters fundamentally affects the switching mechanisms through several physical processes: Joule heating distribution throughout the device, dynamics of filament formation and rupture, oxygen vacancy migration patterns, and the development of local temperature gradients. These findings emphasize that achieving optimal device performance requires precise tuning of both electrical and thermal conductivities to maintain uniform switching characteristics. Understanding these relationships provides crucial insights for material engineering and device optimization, particularly for applications requiring high reliability and consistency in resistive switching behavior.

**Conclusion**

This research provides comprehensive insights into how electrical and thermal conductivities fundamentally influence the operation of oxide-based RRAM devices. Through detailed physical modeling and simulation, we demonstrate that the interplay between transport properties critically influences key device characteristics such as forming voltage, resistance ratio, and uniformity of resistance states. Our analysis reveals several important findings. First, the forming voltage exhibits a clear dependence on both conductivities: higher electrical conductivity promotes oxygen vacancy generation and reduces forming voltage, while higher thermal conductivity leads to increased forming voltage through enhanced heat dissipation. Second, we identify an optimal region for achieving high resistance ratios at moderate conductivity values, where balanced Joule heating and heat dissipation enable both efficient filament formation and complete rupture. Our finding shows two distinct pathways for achieving optimal HRS uniformity, each governed by different physical mechanisms: one at moderate electrical and low thermal conductivity, and another at higher values of both conductivities. For LRS, high uniformity can be maintained when either conductivity is low, offering greater flexibility in material design. These findings provide crucial guidelines for oxide material selection in RRAM devices, suggesting that materials with carefully tuned transport properties can significantly enhance device performance and reliability. The existence of multiple pathways to achieve optimal uniformity offers flexibility in material engineering while highlighting the importance of balanced transport properties. This understanding paves the way for designing more efficient and reliable RRAM devices for large-scale integration in next-generation memory and neuromorphic computing applications.